\documentclass{PoS}
\usepackage{multirow}

\title{First combined studies on Lorentz Invariance Violation from observations of astrophysical sources}

\ShortTitle{LIV Combined studies}

\author{\speaker{Leyre Nogu\'es}$^{1}$, Tony T.Y. Lin$^{2}$, Cedric Perennes$^{3}$, Alasdair E. Gent$^{4}$, Julien Bolmont$^{3}$, Markus Gaug$^{5}$, Agnieszka Jacholkowska$^{3}$, Manel Martinez$^{1}$, A.Nepomuk Otte$^{4}$, Robert M. Wagner$^{6}$, John E. Ward$^{1}$, Benjamin Zitzer$^{2}$  for the LIV Consortium\\
       $^{1}$Institut de F\'isica d'Altes Energies (IFAE), The Barcelona Institute of Science and Technology (BIST), Barcelona, Spain\\
       $^{2}$McGill University, Montreal, Canada\\
       $^{3}$Sorbonne Universit\'es, UPMC Universit\'es Paris 06, Universit\'es Paris Diderot, Sorbonne Paris Cit\'e, CNRS, Laboratoire de Physique Nucl\'eaire et de Hautes Energies (LPNHE), 4 place Jussieu, F-75252, Paris Cedex 5, France\\
       $^{3}$McGill University, Montreal, Canada\\
       $^{4}$Georgia Tech, Atlanta, U.S.A.\\
       $^{5}$Universitat Aut\`onoma de Barcelona, Barcelona, Spain\\
       $^{6}$Stockholms universitet, Sweden\\
       E-mail: \email{lnogues@ifae.es}}

\abstract{Imaging Atmospheric Cherenkov Telescopes study the highest energy (up to tens of TeV) photon emission coming from nearby and distant astrophysical sources, thus providing valuable results from searches for Lorentz Invariance Violation (LIV) effects. Highly variable, energetic and distant sources such as Pulsars and AGNs are the best targets for the Time-of-Flight LIV studies. However, the limited number of observations of AGN flares or of high-energy pulsed emission greatly restricts the potential of such studies, especially any potential LIV effects as a function of redshift.

To address these issues, an inter-experiment working group has been established by the three major collaborations taking data with Imaging Atmospheric Cherenkov Telescopes  (H.E.S.S., MAGIC and VERITAS) with the aim to increase sensitivity to any effects of LIV, together with an improved control of systematic uncertainties, by sharing data samples and developing joint analysis methods. This will allow an increase in the number of available sources and to perform a sensitive search for redshift dependencies.

This presentation reviews the first combined maximum likelihood method analyses using simulations of published source observations done in the past with H.E.S.S., MAGIC and VERITAS. The results from analyses based on combined maximum likelihood methods, the strategies to deal with data from different types of sources and instruments, as well as future plans will be presented.}

\FullConference{35th International Cosmic Ray Conference --- ICRC2017\\
		10--20 July, 2017\\
		Bexco, Busan, Korea}

\begin{document}

\section{Introduction} \label{sec1}
The constancy of the speed of light in vacuum is a cornerstone of Einstein's theories of Relativity. However, the common theoretical framework for Gravity and Quantum Mechanics, still a work in progress, may induce quantum effects in the space-time structure at Planck scale (e.g. space-time foam proposed in \cite{Wheerler}) which would result in an energy dependent speed of light. This would imply that the Lorentz Invariance is not an exact symmetry of
vacuum. Various ways may lead to Lorentz Invariance Violation (LIV): String Theories, Loop Quantum Gravity, non-commutative geometry or modified Special Relativity \cite{Rovelli,AmelinoCamelia:2000, Ellis}. One way to test LIV is to use measurements of the energy dependent time-lags in the light-curves of the very high-energy (VHE) photons coming from distant astrophysical sources \cite{AmelinoCamelia:1997}.  This new window on Quantum Gravity (QG) effects, could allow to
discard an important number of theoretical models. The time-of-flight studies with photons aim at constraining the linear or quadratic terms in the modified dispersion relations connecting energy and momentum of the photon:
\begin{equation}\label{eq:disp}
E^{2}\simeq p^{2}c^{2}\times\left[1-\sum_{n=1}^{\infty}\pm\left(\frac{E}{E_{QGn}}\right)^{n}\right],
\end{equation}
where $E_{QGn}$ is the Quantum Gravity scale, the $\pm$ sign refers to subluminal or superluminal corrections to the speed of light and $n$ takes value of 1 or 2.
Significant efforts during last fifteen years based on observations of several GRBs, AGN flares or pulsars at high energies allow already to constrain linear or quadratic terms in \ref{eq:disp} (See for review \cite{Horns:2016soz} and references in). For the linear case, limits on $E_{QG}$ even attains the Planck energy scale \cite{GRB}. The constraints on the quadratic term stay several orders of magnitude below the Planck scale and will remain a challenge for future studies. However, it has to be noted that the use of photons as messengers reach some limitations due to energy-dependent time-lags
produced during photon emission in the astrophysical sources. Another restriction comes from
the so-called gamma-ray horizon which limits the energy range of detected photons due to absorption by the Extragalactic Background Light (EBL).

To improve the sensitivity of LIV studies and to further extend constraints on various models,
a combination of results from different types of sources provided by the three major Cherenkov Telescope experiments (H.E.S.S., MAGIC and VERITAS) has been studied. The proposed procedure described in the following improves statistical power of published studies and tends to minimize systematic uncertainties related to each individual source measurement. Moreover, the predicted linear dependence of the LIV effects on the source redshift allows to discard hypothesis of the source emission intrinsic time-lags which are in principle redshift independent.

In the following paper the likelihood method for the time-lag determination and the procedure
of the source combination are presented and discussed followed by the results obtained from simulations of the existing published data for AGNs and Pulsars. The combined results from three AGNs and one Pulsar are compared with those obtained for each individual source. Finally, the scientific impact of these first results is presented and future plans and prospects are discussed.

\section{Methodology}\label{sec2}
All observatories considered for combination in this work use the Maximum Likelihood (ML) method for the extractions of the LIV limits. Compared with alternative methods developed in the literature, the ML allows an optimal use of the information contained in the data and gives a measurement of the probability that allows a rather straightforward combination of the results from the different observatories. The ML method conceptually relies on the definition of the Probability Distribution Function (PDF) that describes the probability of a gamma-ray being observed with a given energy and arrival time, assuming a certain energy-dependent delay function and taking into account the instrument effects in the measurement. The first time this approach was proposed was in \cite{Manel}, where the event PDF formula for flaring sources reads

\begin{equation}\label{eq:pdf}
\frac{dP}{dEdt}=N\int_0^{\infty}\Gamma(E_s)C(E_s,t)G(E-E_s,\sigma_E(E_s))F_s(t-D(E_s, E_{QGn},z))dE_s,
\end{equation}
where $\Gamma(E_s)$ is the photon energy distribution at the source, $C(E_s,t)$ is the collection area, $G(E-E_s,\sigma_E(E_s))$ is the instrument energy smearing, $F_s(t)$ is the emission distribution time at the source and $D(E_s, E_{QGn},z)$ is the energy-dependent propagation delay. The likelihood function ($L$) is built with the PDF of every event and has at least one parameter, the estimator, related to $E_{QGn}$. The aim of the method is to find the value of the estimator that maximizes the likelihood.

In practice, the concept behind the above formula has been applied in different manners by the different observatories and the different source types \cite{Mrk501, PKS2155, PG:2015ixa, Crab, GRB}. For instance some observatories use unbinned data while others use binned data. Also, sometimes the ML fit is multi-parametric, where some quantities are treated as nuisance parameters and profiled to propagate their uncertainty, whereas in others is uni-parametric combined with Monte Carlo simulations to propagate the uncertainties in the possible additional parameters. On the one hand some observatories deal with sources that have flares (AGNs) and the gamma-ray arrival time is used while, on the other hand, others have periodic emissions (pulsars) and the gamma-ray arrival phase is used instead.

Nevertheless, at the end all observatories do deliver likelihood functions for the different sources, with a common LIV parameter or estimator, that can be combined into a single likelihood $L_{Comb}$ allowing a joint parameter estimation

\begin{equation}\label{eq:combLik}
L_{Comb}(\lambda) = \prod_{i=1}^{Nsource}L_i(\lambda)\,\,\,\longrightarrow \,\, -2log(L_{Comb}(\lambda)) = -2\sum_{i=1}^{Nsource} log(L_i(\lambda)),
\end{equation}
where $\lambda$ is the LIV parameter. To combine different sources from different experiments, the common LIV parameter must be redshift independent.

Typically each likelihood function has a parabolic shape in logarithmic scale close to the minimum and therefore the combination of the results from the different sources and observatories consist in combining in logarithmic scale the sum of the parabolas of each measurement, as shown in Formula \ref{eq:combLik}. Looking for a maximum in the Likelihood is equivalent to looking for a minimum in a negative logarithmic scale.

Once the measurements are combined into a single parabola as a function of the LIV parameter, Confidence Levels (CLs) for either a measurement, if the parabola minimum is significantly different from the non-LIV effect hypothesis, or a single-sided CLs can be easily extracted. In this work, 1-sided 95\% CLs are extracted from the crossing point between the corresponding curve and the $-2log(L_{Comb}(\lambda)) = 2.71$ line.

\section{Data Simulations}\label{sec3}
The individual sources combined in this work are three AGN flares -- Mrk 501 2005 flare detected by MAGIC \cite{Mrk501}, PG 1553+113 2012 flare detected by H.E.S.S \cite{PG:2015ixa} and PKS 2155-304 2006 flare detected by H.E.S.S \cite{PKS2155} -- and VHE radiation from the Crab Pulsar detected by VERITAS \cite{Crab}. The individual simulation settings are summarized in Table \ref{tab:simulation}.
The first two terms, linear and quadratic in energy in formula \ref{eq:disp}, were considered for the simulation and later analysis.

For testing of LIV, and hence the corresponding QG models, we generate simulated data sets constructed from parametrization of published observational data from the sources mentioned above. 

\begin{figure}
\caption{Simulated PKS 2155-304 data. Left - time distribution and right - energy distribution (Upper panel, true energy, lower panel, measured energy).}\label{fig:SIM_PKS}
\centering
\includegraphics[scale=0.3]{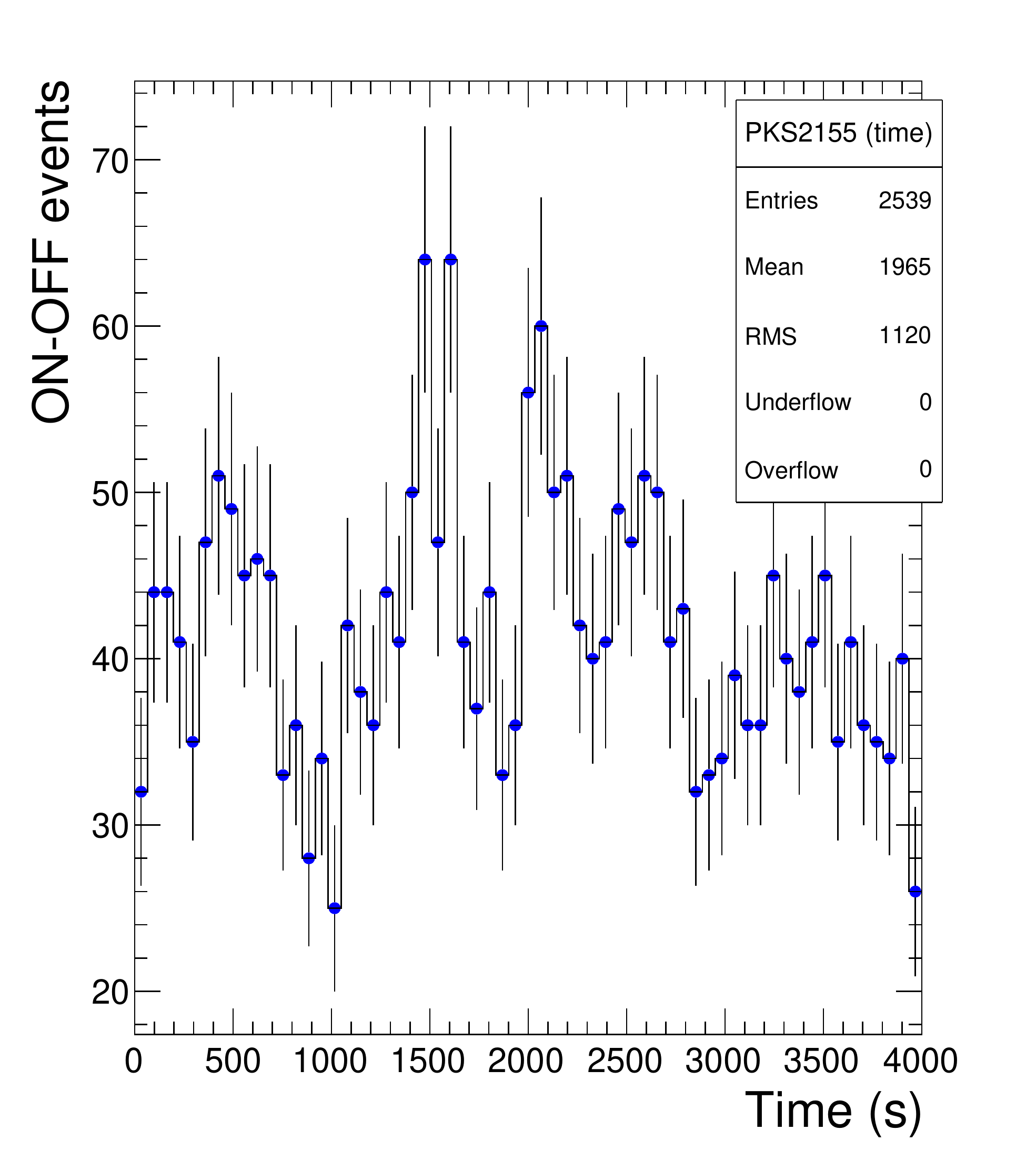}
\includegraphics[scale=0.3]{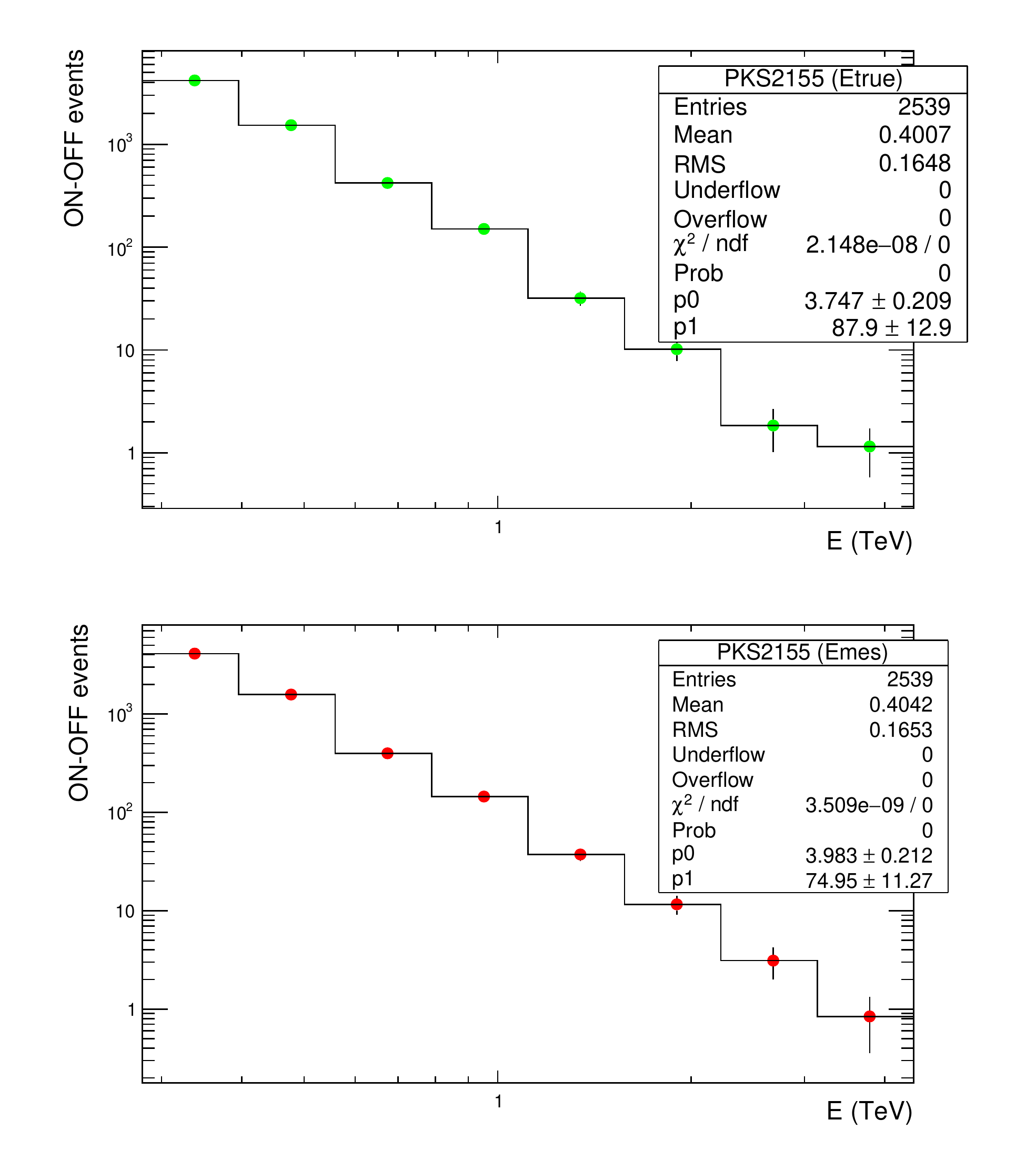}
\end{figure}

\begin{table}
\resizebox{\textwidth}{!}{%
%\begin{center}
    \begin{tabular}{ |c|c|c|c|c|c|c|} 
     \hline
     \multicolumn{7}{|c|} {Simulation settings} \\ \hline
     Source & Energy Range (TeV) & Time Range & Spectral shape & Lightcurve & Energy resolution & Number of events \\ \hline
     \textbf{PG 1553+113} & 0.4 - 0.8 & 0 - 8000 s & PL (Index = 4.8) & Double Gauss & 10\% & 180\\ \hline
     \textbf{Mrk501} & 0.25 - 11 & 0 - 1531 s & PL (Index = 2.2) & Simple Gauss & 22\% & 800\\ \hline
     \textbf{PKS 2155-304} & 0.28 - 4 & 0 - 4000 s & PL (Index = 3.46) & 5 Asymmetric Gauss & 10\% & 2800\\ \hline
     \textbf{Crab} & 0.12 - 7 & 0 - 1 phase & PL (Index = 3 for signal, 3.5 for bkg) & Double Gauss + Baseline & 10\% & 860000 \\ \hline
     \end{tabular}}
      \caption{Simulation settings for the individual sources.}
      \label{tab:simulation}
%\end{center}
\end{table}

Event true energies and arrival times are generated from the parametrized spectra and lightcurves for each source. The number of simulated events also follows the real data. The LIV time-lag effect ($\Delta t \propto E^n$) is added, as required for linear or quadratic model. Instrument Response Functions (IRFs) are used to model detection probability and reconstructed energy for each event. In this paper, the results for a zero time-lag are presented. Only signal events are considered. For each source, 990 measurements are simulated to be used for analysis. 

The Monte Carlo (MC) simulation data is analyzed to show that the spectra and lightcurves are compatible with the original data. As an example, Figure \ref{fig:SIM_PKS} shows the simulated time and energy distributions for the  PKS 2155-304 flare.

\section{Results on LIV and QG limits}\label{sec4}
As discussed in section \ref{sec1}, the expected energy-dependent time lag for photons $\Delta t/\Delta E^n$ can be related to $E_{QGn}$ as

\begin{equation}
\label{eq:energy}
\frac{\Delta t_n}{E_h^n - E_l^n} \simeq s_{\pm}\,\frac{n+1}{2\,H_0}\,\frac{1}{E_{QGn}} \int_0^z \frac{(1+z')^n}{\sqrt{\Omega_m\,(1+z')^3 + \Omega_\Lambda}}\,dz' = s_\pm\,\frac{n+1}{2\,H_0}\,\frac{1}{E_{QGn}^n} \kappa(z),
\end{equation}

being $\Omega_m$ and $\Omega_\Lambda$ the standard cosmological parameters.

 Limits on $\Delta t/\Delta E^n$ provided an estimation on the $E_{QGn}$ at 1-sided 95\% CLs. The Formula \ref{eq:energy} taking into account the expansion of the Universe (parameters $\Omega_m$ and $\Omega_\Lambda$ in \ref{eq:energy}) was used for the limit calculations for n equal 1 or 2, parameter $\kappa_q(z)$ being referred as $\kappa_l(z)$ for the linear and $\kappa_q(z)$ for the quadratic case. 
 
 The use of $\lambda$ as a fit parameter, defined as
 
 \begin{equation}\label{eq:lambda}
\lambda = \frac{\Delta t_n}{\Delta E^n \kappa(z)} =\frac{1}{E_{QGn}H_{0}},
\end{equation}

 allows a simultaneous analysis of sources with different redshifts. The study of precision on parameter $\lambda$ with MC representative simulations of present published data corresponding to AGN flares and pulsed emission of the Crab Pulsar is the main aim of the presented analysis.

\begin{figure}[h]
\caption{Distributions of the best $\lambda$ values with source combinations, left: linear case, right: quadratic case. The curves are results of Gaussian fits.}
\label{fig:sources_combined}
\centering
\includegraphics[width= 8cm, height = 5cm]{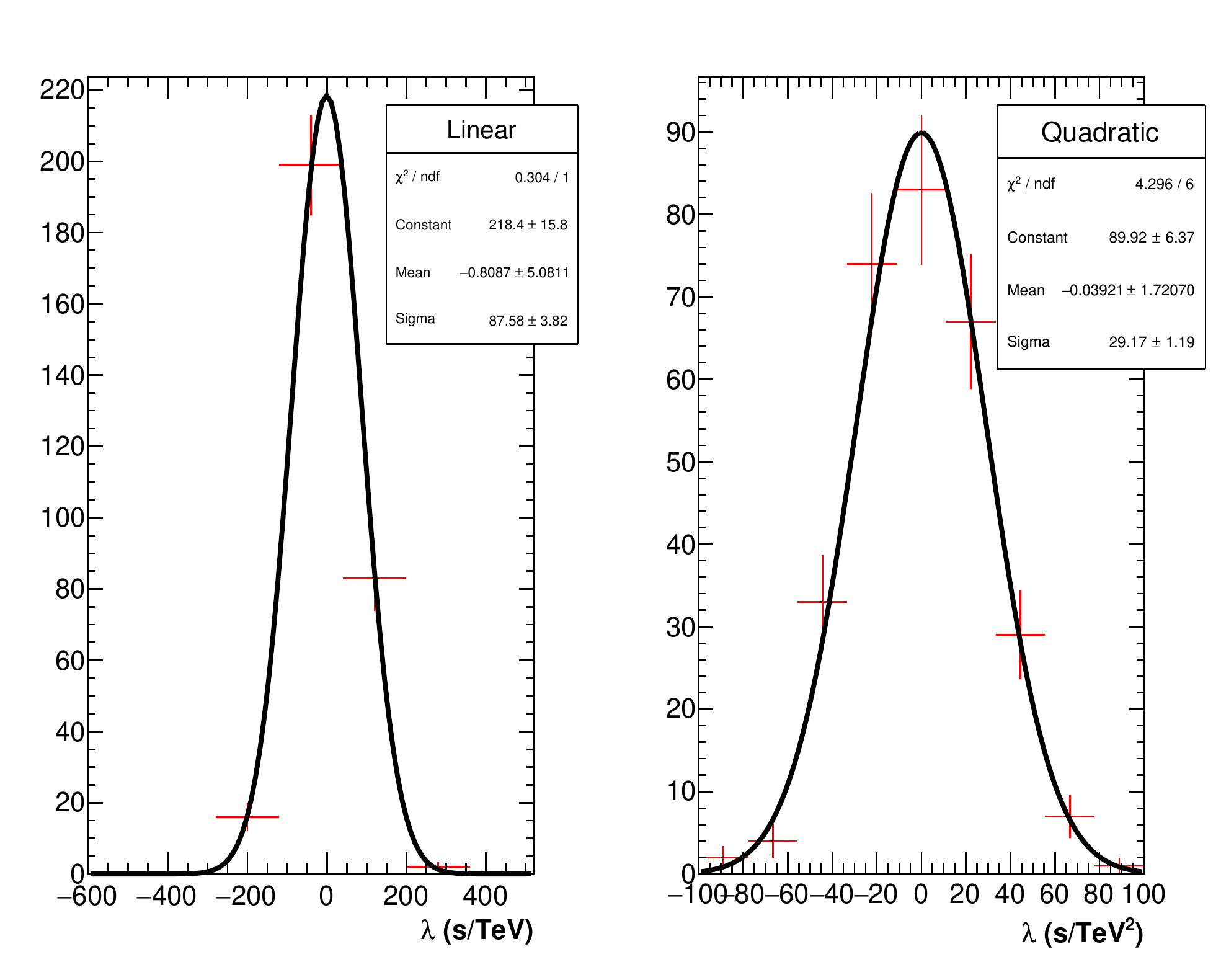}
\end{figure}

The results presented in this section  were obtained with the analysis based on the combined ML method discussed in section \ref{sec2}. To obtain first the estimation of attained precision, no initial time-lag has been injected in the simulated data samples. The simulations of events for different sources followed the procedure described in section \ref{sec4}. The analysis of each source as well as their combination provided the best fit value of the parameter $\lambda$, as well as  1-sided $95\%$ CLs. The distributions of $\lambda$ and CLs were built with 990 realizations allowing to evaluate the statistical probability to obtain a given result. As no systematic effects have been introduced in the likelihood fits, both $\lambda$ and CLs present Gaussian behavior as seen in Figure \ref{fig:sources_combined}. Later, the systematic effect contribution was globally estimated and added in quadrature when calculating the $E_{QGn}$ limits.

\begin{figure}[h]
\caption{Mean best $\lambda$  values and standard deviations (x-axis) for each source and combination, as a function of the kappa parameter (y-axis). Left - linear case, right - quadratic case. The values were obtained from Gaussian fits of the best value lambda distributions.}
\label{fig:cern}
\centering
\includegraphics[width= 6.5cm, height = 5cm]{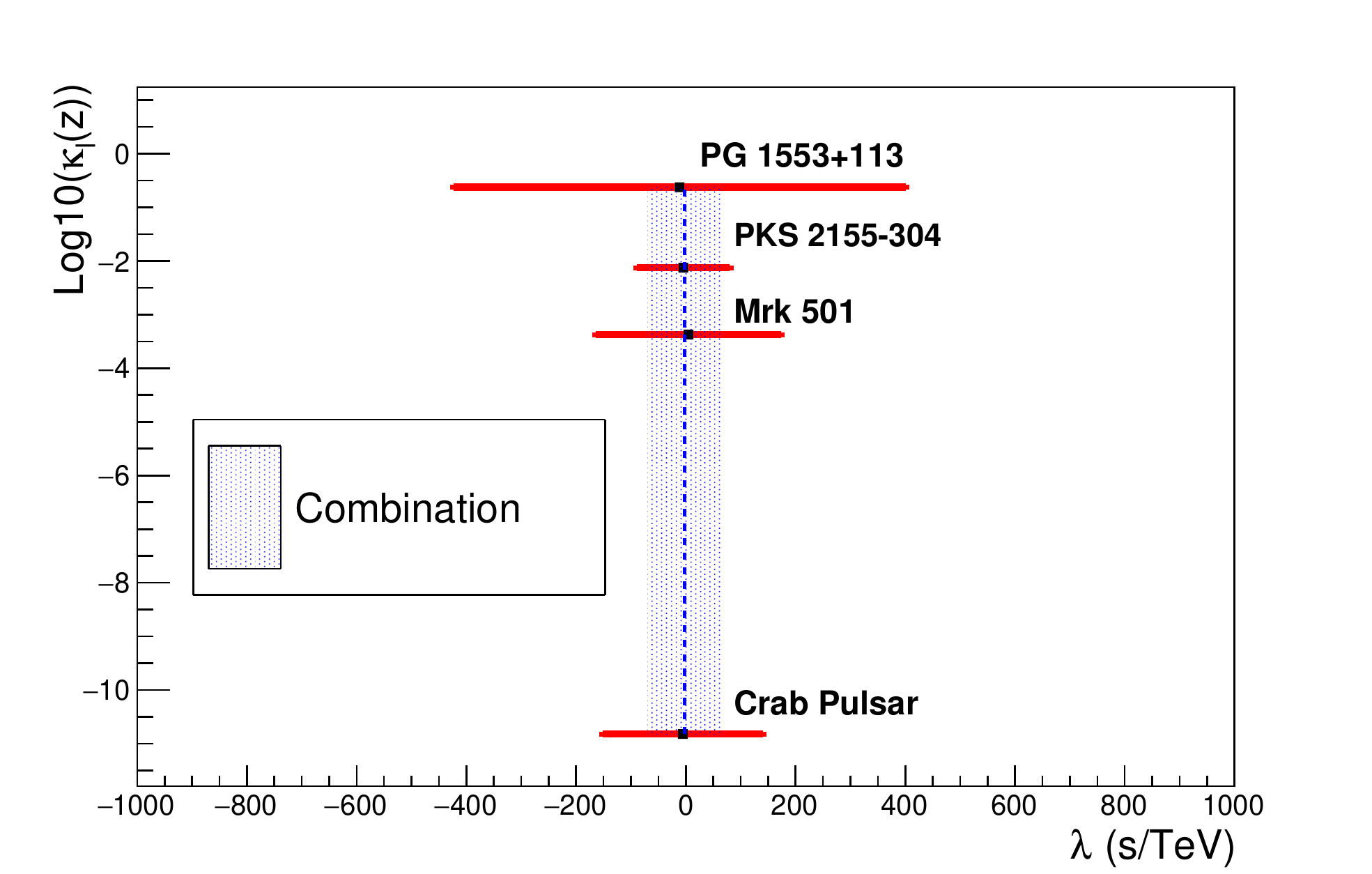}
\includegraphics[width= 6.5cm, height = 5cm]{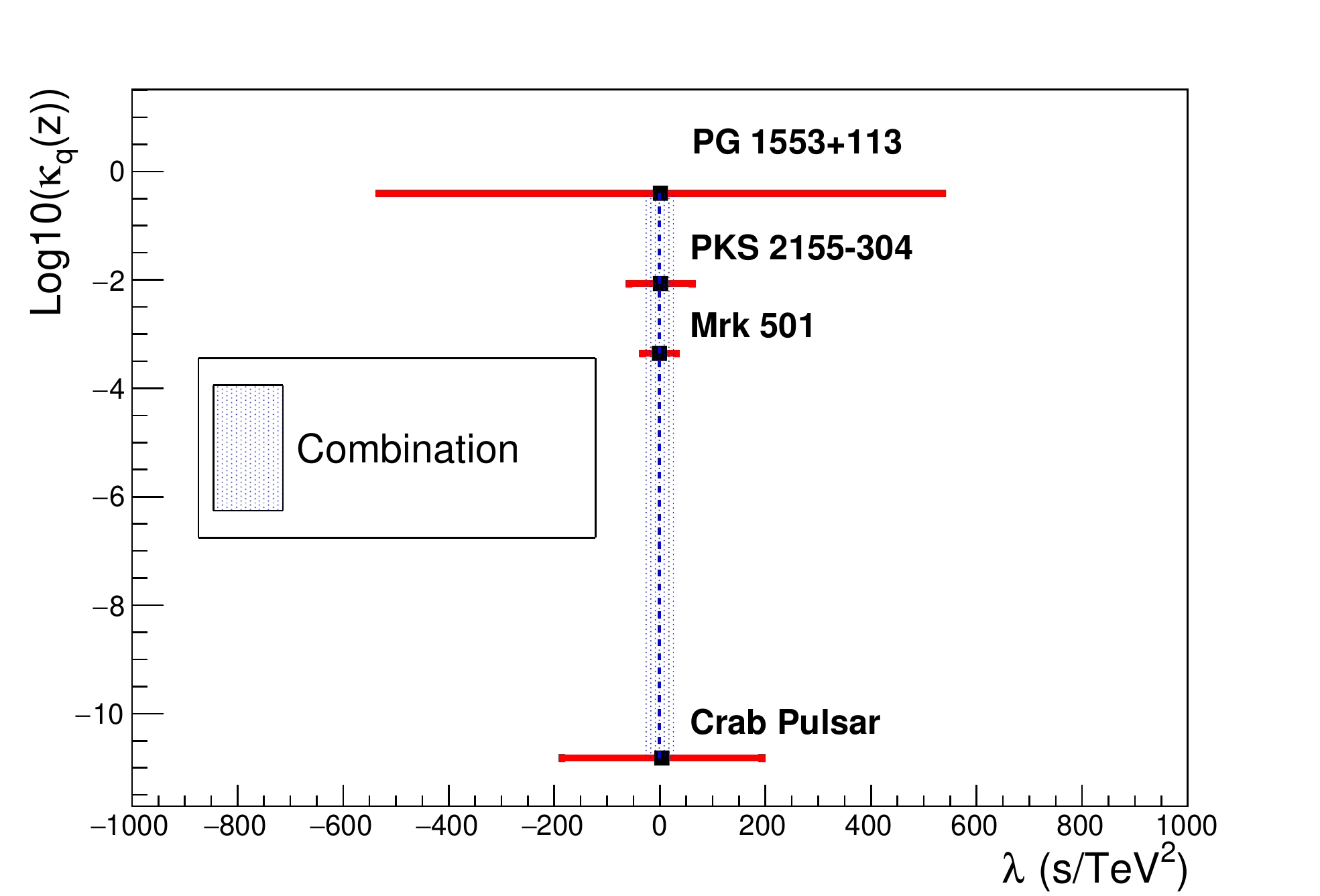}
\end{figure}

 The combination results for both cases and their Gaussian fit are exposed in Figure \ref{fig:sources_combined}. Table \ref{tab:linear} summarizes mean fit values and standard deviations in $\lambda$ for individual and combined cases for linear model and in Table \ref{tab:quadratic} for the quadratic model. A comparison of the results for each source and their combination is shown in Figure \ref{fig:cern}. It should be noted that independent of the source redshift, the reconstructed mean values of the time-delay reproduce well the initial value of time-lag equal to zero within 1 $\sigma$ deviation. Thus no systematic shift is introduced by the method in use.

\begin{table}
\begin{center}
    \begin{tabular}{ |c|c|c|c|c|c|} 
      \hline
     \textbf{Parameter} & \textbf{PKS 2155} & \textbf{Mrk 501} & \textbf{PG 1553}  & \textbf{Crab} & \textbf{Combination} \\ \hline
     $\lambda_{best}$ ($s/TeV$) & -4.5$\pm$2.6 & 4.9$\pm$5.6 & -11.3$\pm$13.4 & -5.4$\pm$4.7 & -2.37$\pm$2.2\\ \hline
     $1\sigma\,\,CL $ ($s/TeV$) & 84.6$\pm$2.1 & 168.6$\pm$4.4 & 412.0$\pm$9.7 & 146.0$\pm$3.8 & 67.6$\pm$1.6\\ \hline
     $\lambda_{LL}$ ($s/TeV$) & -154.9 & -296.6 & -687.7 & -254.2 & -118.2\\ \hline
     $ RMS_{LL}$ ($s/TeV$) & 88.8 & 169.9 & 414.5 & 150.4 & 67.52\\ \hline
     $\lambda_{UL}$ ($s/TeV$) & 142.5 & 299.5 & 658.6 & 244.7 & 117.8\\ \hline
     $ RMS_{UL}$ ($s/TeV$) & 83.72 & 171.6 & 421.4 & 151.3 & 66.1\\ \hline
     \end{tabular}
    \caption{Linear case: best $\lambda$ values and 1-s 95\% CL Upper Limits on  $\lambda$ for each source and combination. The standard deviations and RMS values for the limits are also shown}  \label{tab:linear}
\end{center}
\end{table}

\begin{table}
\begin{center}
    \begin{tabular}{ |c|c|c|c|c|c|} 
     \hline
     \textbf{Parameter} & \textbf{PKS 2155} & \textbf{Mrk 501} & \textbf{PG 1553} & \textbf{Crab} & \textbf{Combination} \\ \hline
     $\lambda_{best}$ ($s/TeV^{2}$) & 1.3$\pm$1.9 &  -0.8$\pm$1.1 & 1.0$\pm$17.5 & 3.8$\pm$6.4 & -0.6$\pm$0.9\\ \hline
     $1\sigma\,\,CL $ ($s/TeV^{2}$) & 59.8$\pm$1.7 & 31.85$\pm$1.0 & 533.7$\pm$13.2 & 189.5$\pm$5.6 & 26.7$\pm$0.7\\ \hline
     $\lambda_{LL}$ ($s/TeV^{2}$) & -104.4 & -59.2 & -912.1 & -326.6 & -49.5\\ \hline
     $ RMS_{LL}$ ($s/TeV^{2}$) & 69.2 & 33.2 & 542.1 & 351.0 & 28.9 \\ \hline
     $\lambda_{UL}$ ($s/TeV^{2}$) & 100.0 & 56.8 & 921.1 & 354.2 & 48.1 \\ \hline
     $ RMS_{UL}$ ($s/TeV^{2}$) & 67.9 & 34.1 & 554.2 & 355.0 & 28.0 \\ \hline
     \end{tabular}
      \caption{Quadratic case: best $\lambda$ values and 1-s 95\% CLs Upper Limits on Lambda for each source and combination. The standard deviations and RMS values for the limits are also shown.}\label{tab:quadratic} 
\end{center}
\end{table}

\begin{figure}[h]
\caption{Lower limits on $E_{QGn}$ as a function of parameter kappa (Redshift). Sub-luminal case considered. Left - linear case and right - quadratic case. $E_{Plank}=1.22\cdot10^{19}$.}\label{fig:Elim}
\centering
\includegraphics[scale=0.3]{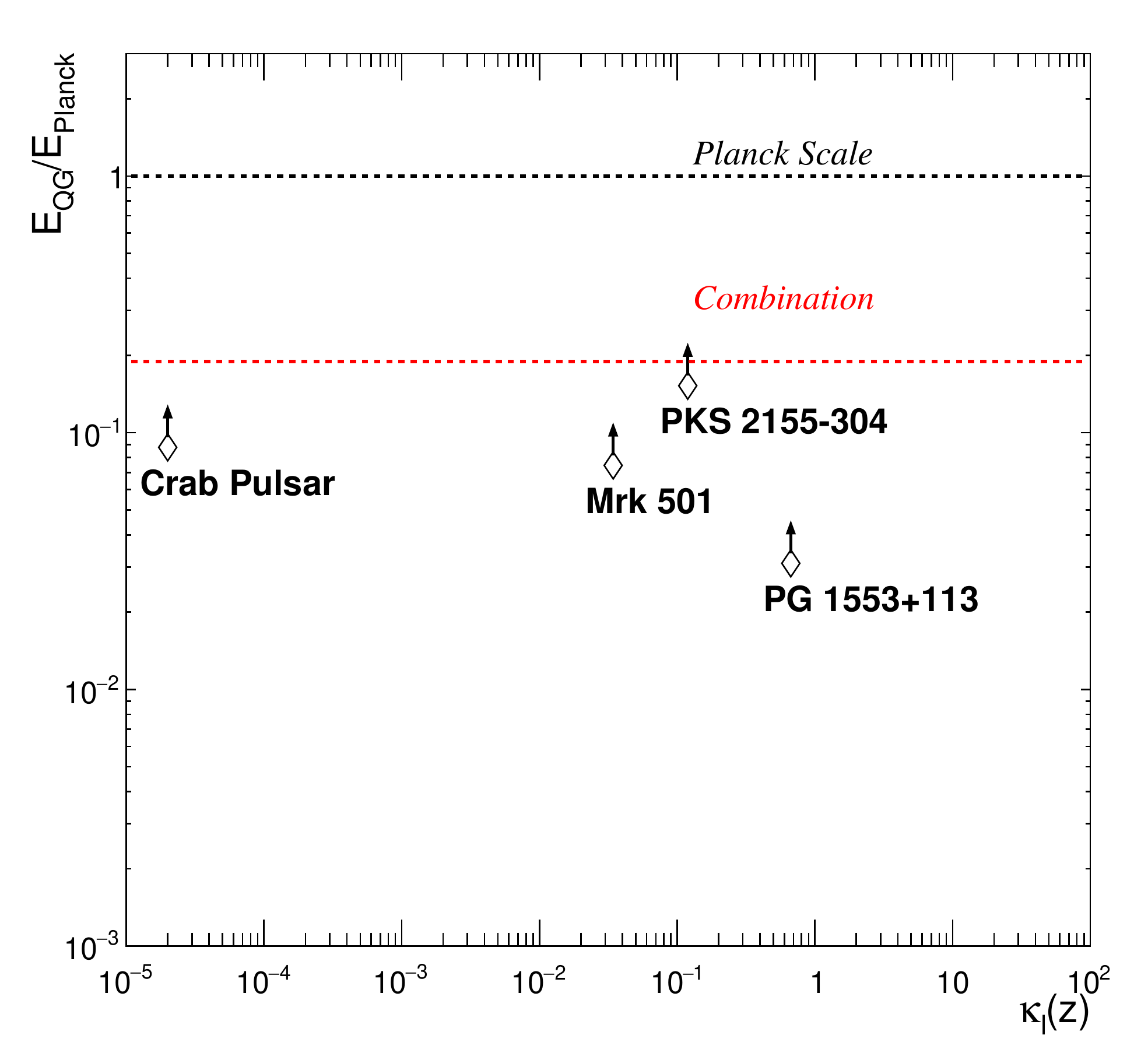}
\includegraphics[scale=0.3]{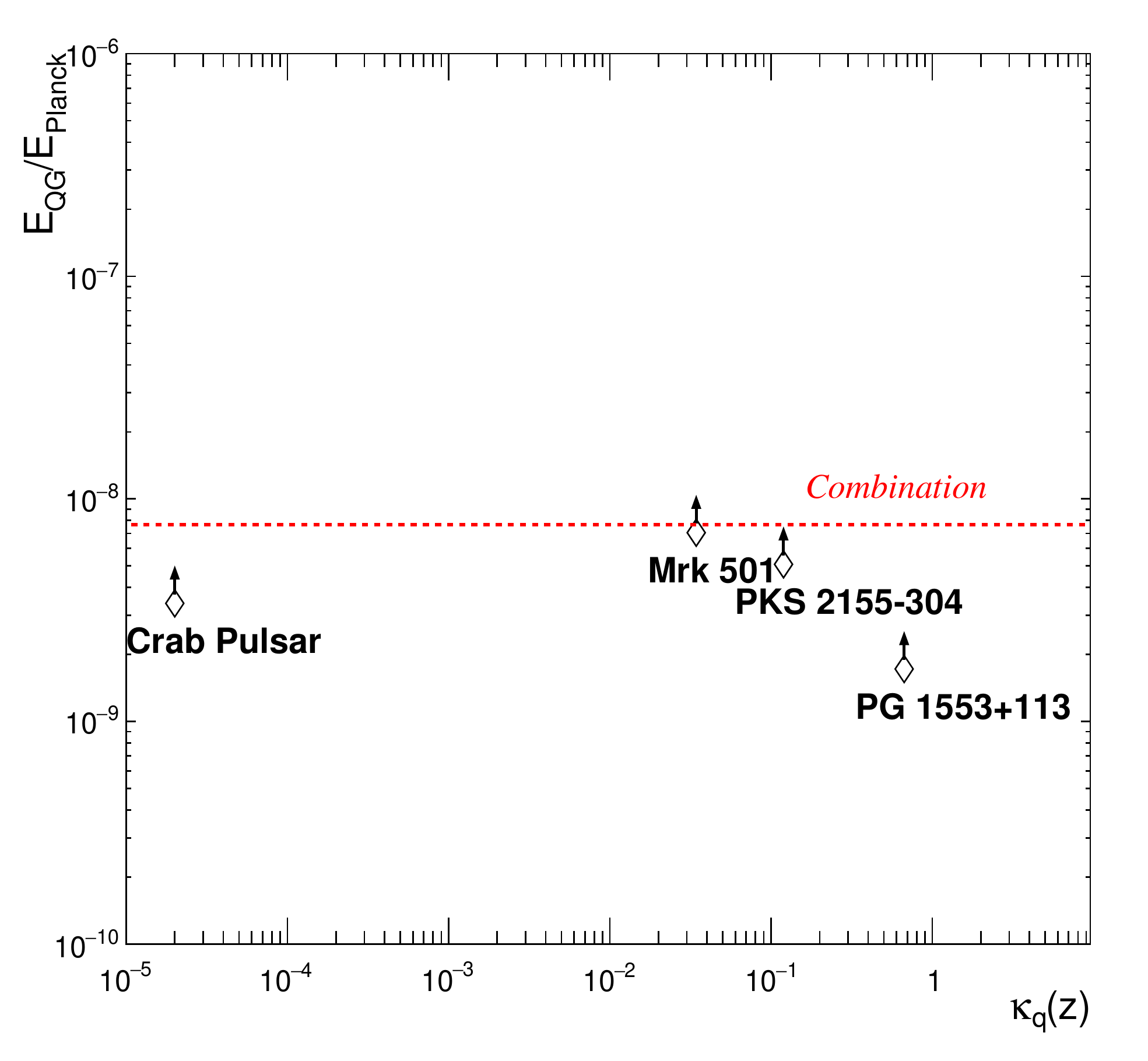}
\end{figure}

The limits on the $E_{QGn}$ energy scale are collected in Table \ref{tab:energy} and graphically represented in Figure \ref{fig:Elim}. These limits were computed considering both statistical and systematic error; the systematic uncertainty is conservatively take to be equal to the statistical error in this study. As the main conclusion, the limits improve with combination procedure already on the level of $\lambda$ parameter. Still one can observe that in linear case the results are strongly dominated by the PKS 2155-304 limit, less relevant for the quadratic case, where Mrk 501 provides already an outstanding result. In the quadratic case a 26\% improvement is obtained with combination and of 10\% in linear case respectively. The source PG 1553+113 provides important results at very-high-redshift even if contributing on a lower level to the combination but extending redshift range closer to those found in GRBs \cite{GRB}. Crab Pulsar also contributes in the redshift range but, specially, by increasing importantly the number of events. In the future, more AGN flares will be added to this study leading to a larger extension of the redshift range.
\begin{table}
\begin{center}
    \begin{tabular}{ |c|c|c|c|} 
     \hline
     Source & $E_{QG\_linear}(10^{18} GeV)$ & $E_{QG\_Quadratic}(10^{10} GeV)$ & Redshift \\ \hline
     \textbf{PKS 2155} & 1.86 & 6.20 & 0.116\\ \hline
     \textbf{Mrk 501} & 0.91 & 8.57 & 0.034\\ \hline
     \textbf{PG 1553} & 0.38 & 2.08 & 0.5\\ \hline
     \textbf{Crab} & 1.07 & 4.14 & 2kpc \\ \hline
     \textbf{Combination} & 2.31 & 9.34 & - \\ \hline
     \end{tabular}
      \caption{1-s 95\% CL Upper Limits on QG energy scale for linear and quadratic case, for each source and combination. }
      \label{tab:energy}
\end{center}
\end{table}

Further conclusions to be drawn from this study is a better determination of the results when taking the mean values of the upper and lower limits as shown with RMS values in Tables 1 and 2 which decrease when combined. This type of effect is also expected when introducing the systematic effects with nuisance parameters directly into the likelihood fit which is foreseen in future studies and applied for the real data analyses.

\section{Conclusions and prospects}\label{sec5}
We have shown here the first results of an inter-experiment working group consisting of members of the H.E.S.S., MAGIC and VERITAS collaborations. The joint ML allows for not only one individual source but sources from several different instruments to be combined in a relatively straightforward manner. Simulations generated from inputs from published source observations are
used to show that a more sensitive search for LIV through time-of-flight measurements can be performed than any one individual search on a single target or instrument.

In the future of these studies we plan to investigate the effects of systematic effects on the measurement and implement them as nuisance parameters directly into the likelihood. The target lists and event lists from each of the collaborations are being finalized. The benefits of the methods presented here can be utilized for all existing and future gamma-ray experiments, including the upcoming Cherenkov Telescope Array (CTA) \cite{CTA}.

\bibliographystyle{JHEP}
\bibliography{references}

\providecommand{\href}[2]{#2}\begingroup\raggedright\begin{thebibliography}{10}

\bibitem{Wheerler}
J.~Wheeler, \emph{Relativity, Groups and Topology}.
\newblock Gordon and Breach, New York, 1964.

\bibitem{Rovelli}
C.~Rovelli, \emph{{The Century of the incomplete revolution: Searching for
  general relativistic quantum field theory}},
  \href{http://dx.doi.org/10.1063/1.533327}{\emph{J. Math. Phys.} {\bfseries
  41} (2000) 3776--3800},
  [\href{https://arxiv.org/abs/hep-th/9910131}{{\ttfamily hep-th/9910131}}].

\bibitem{AmelinoCamelia:2000}
G.~Amelino-Camelia, \emph{{Quantum theory's last challenge}},
  \href{http://dx.doi.org/10.1038/35047210, 10.1038/408661a}{\emph{Nature}
  {\bfseries 408} (2000) 661--664},
  [\href{https://arxiv.org/abs/gr-qc/0012049}{{\ttfamily gr-qc/0012049}}].

\bibitem{Ellis}
J.~R. Ellis, N.~E. Mavromatos and D.~V. Nanopoulos, \emph{{A microscopic recoil
  model for light cone fluctuations in quantum gravity}},
  \href{http://dx.doi.org/10.1103/PhysRevD.61.027503}{\emph{Phys. Rev.}
  {\bfseries D61} (2000) 027503},
  [\href{https://arxiv.org/abs/gr-qc/9906029}{{\ttfamily gr-qc/9906029}}].

\bibitem{AmelinoCamelia:1997}
G.~Amelino-Camelia, J.~R. Ellis, N.~E. Mavromatos, D.~V. Nanopoulos and
  S.~Sarkar, \emph{{Tests of quantum gravity from observations of gamma-ray
  bursts}}, \href{http://dx.doi.org/10.1038/31647}{\emph{Nature} {\bfseries
  393} (1998) 763--765},
  [\href{https://arxiv.org/abs/astro-ph/9712103}{{\ttfamily
  astro-ph/9712103}}].

\bibitem{Horns:2016soz}
D.~Horns and A.~Jacholkowska, \emph{{Gamma-rays as probes of the Universe}},
  \href{http://dx.doi.org/10.1016/j.crhy.2016.04.006}{\emph{Comptes Rendus
  Physique} {\bfseries 17} (2016) 632--648},
  [\href{https://arxiv.org/abs/1602.06825}{{\ttfamily 1602.06825}}].

\bibitem{GRB}
V.~Vasileiou, A.~Jacholkowska, F.~Piron, J.~Bolmont, C.~Couturier, J.~Granot
  et~al., \emph{{Constraints on Lorentz Invariance Violation from Fermi-Large
  Area Telescope Observations of Gamma-Ray Bursts}},
  \href{http://dx.doi.org/10.1103/PhysRevD.87.122001}{\emph{Phys. Rev.}
  {\bfseries D87} (2013) 122001},
  [\href{https://arxiv.org/abs/1305.3463}{{\ttfamily 1305.3463}}].

\bibitem{Manel}
M.~Martinez and M.~Errando, \emph{A new approach to study energy-dependent
  arrival delays on photons from astrophysical sources},
  \href{http://dx.doi.org/https://doi.org/10.1016/j.astropartphys.2009.01.005}{\emph{Astroparticle
  Physics} {\bfseries 31} (2009) 226 -- 232}.

\bibitem{Mrk501}
J.~Albert et~al., \emph{{Variable VHE gamma-ray emission from Markarian 501}},
  \href{http://dx.doi.org/10.1086/521382}{\emph{Astrophys. J.} {\bfseries 669}
  (2007) 862--883}, [\href{https://arxiv.org/abs/astro-ph/0702008}{{\ttfamily
  astro-ph/0702008}}].

\bibitem{PKS2155}
F.~Aharonian, \emph{{An Exceptional Very High Energy Gamma-Ray Flare of PKS
  2155-304}}, \href{http://dx.doi.org/10.1086/520635}{\emph{Astrophys. J.}
  {\bfseries 664} (2007) L71--L78},
  [\href{https://arxiv.org/abs/0706.0797}{{\ttfamily 0706.0797}}].

\bibitem{PG:2015ixa}
{\scshape H.E.S.S.} collaboration, A.~Abramowski et~al., \emph{{The 2012 flare
  of PG 1553+113 seen with H.E.S.S. and Fermi-LAT}},
  \href{http://dx.doi.org/10.1088/0004-637X/802/1/65}{\emph{Astrophys. J.}
  {\bfseries 802} (2015) 65},
  [\href{https://arxiv.org/abs/1501.05087}{{\ttfamily 1501.05087}}].

\bibitem{Crab}
{\scshape VERITAS} collaboration, T.~Nguyen, \emph{{Updated Results from
  VERITAS on the Crab Pulsar}}, {\emph{PoS} {\bfseries ICRC2015} (2016) 828},
  [\href{https://arxiv.org/abs/1508.07268}{{\ttfamily 1508.07268}}].

\bibitem{CTA}
M.~{Actis}, G.~{Agnetta}, F.~{Aharonian}, A.~{Akhperjanian}, J.~{Aleksi{\'c}},
  E.~{Aliu} et~al., \emph{{Design concepts for the Cherenkov Telescope Array
  CTA: an advanced facility for ground-based high-energy gamma-ray astronomy}},
  \href{http://dx.doi.org/10.1007/s10686-011-9247-0}{\emph{Experimental
  Astronomy} {\bfseries 32} (Dec., 2011) 193--316},
  [\href{https://arxiv.org/abs/1008.3703}{{\ttfamily 1008.3703}}].

\end{thebibliography}\endgroup



\providecommand{\href}[2]{#2}\begingroup\raggedright\endgroup

\end{document}